\documentclass[]{imag-ms-template}

\usepackage{amsmath,amssymb}
\usepackage{hyphenat}

\title{TractSpLearn: Specialized Shared-Manifold Learning for Individualized Detection of Subtle White Matter Abnormalities in Mild Traumatic Brain Injury} 

\author{Jiqing Huang $^{1}$, Ali Al-Husseini $^{2\sharp}$, Yi Chen $^{3,4\sharp}$, Anna Gard $^{2}$, \\
Laurent Lamalle $^{1}$,  Mohamed Ali Bahri $^{1}$,  Markus Nilsson $^{5}$, \\
Niklas Marklund $^{2\dagger}$, Christophe Phillips $^{1\dagger}$, Evgenios N. Kornaropoulos $^{1,5,6\ast\dagger}$\\
{\small $^{1}$ GIGA -- CRC Human Imaging, University of Liège, Belgium}\\
{\small $^{2}$ Department of Clinical Sciences Lund, Neurosurgery, Lund university, Skane University Hospital, Sweden}\\
{\small $^{3}$ AMIIC Lab, Guizhou University, China}\\
{\small $^{4}$ D-Lab, Maastricht University, Netherlands}\\
{\small $^{5}$ Department of Clinical Sciences Lund, Diagnostic Radiology, Medical Faculty, Lund University, Sweden}\\
{\small $^{6}$ Aix-Marseille Univ, CNRS, CRMBM, Marseille, France}\\
{\small $^\ast$ Correspondence: ekornaropoulos@uliege.be, niklas.marklund@med.lu.se}\\
{\small $^\sharp$ Equal contribution as second authors}
{\small $^\dagger$ Equal contribution as last authors}
}
\begin{document} 
\maketitle
\keywords{Diffusion MRI, Diffusion Kurtosis Imaging, Fiber Tractography, Manifold Learning, Regression, Traumatic Brain Injury}

\newpage
\begin{abstract} \\
Traumatic brain injury (TBI) often leads to subtle white matter damage that remains undetected on conventional MRI. Diffusion kurtosis imaging (DKI), an extension of diffusion tensor imaging (DTI),  provides complementary information on non-Gaussian water diffusion and is sensitive to complex white-matter microstructure. With the advent of ultra-high-field MRI, the spatial resolution and signal-to-noise ratios (SNR) have been significantly enhanced, enabling more precise visualization of subtle abnormalities. Building on these advances, we developed TractSpLearn, an individualized tract-based learning framework that jointly considers within-group variability and between-group differences. Unlike the original TractLearn framework, which learns a normative manifold exclusively from healthy controls, TractSpLearn incorporates both healthy controls and patients to learn a shared manifold with a healthy-anchored representation and an additional patient-related component. To assess the performance of the proposed method, we compared TractSpLearn with the original TractLearn in three cohorts: (i) healthy controls (HC), (ii) athletes with persistent post-concussive syndromes (PPCS), and (iii) athletes with repeated head injuries (RHI), with abnormalities particularly evident in axial kurtosis (AK) and mean diffusivity (MD). In RHI, TractSpLearn highlighted recurrent abnormalities across patients. In the PPCS cohort, the overall group-level differences were more modest, potentially reflecting both limited statistical power due to the small sample size and partial normalization of white-matter alterations during recovery. Still TractSpLearn identified abnormality evidence in more patients and across more affected tracts than TractLearn. 
\end{abstract}

\newpage
\section{Introduction}
Traumatic brain injury (TBI) refers to brain damage caused by an external mechanical force, such as a forceful bump, blow, or jolt to the head, or skull penetration by an object. Beyond the initial impact, TBI often triggers a cascade of secondary pathophysiological processes.  Patients with TBI may experience persistent cognitive impairments, emotional and behavioral disturbances, and sensory or motor deficits. In severe cases, these consequences can lead to long-term disability, with outcomes varying according to the extent and, particularly, the location of the injury. 

Magnetic resonance imaging (MRI) has been extensively used in TBI to investigate macroscopic abnormalities, such as contusions, hemorrhage, and edema. However, non-hemorrhagic lesions, including diffuse axonal injury, often remain undetectable on conventional MRI, limiting its ability to capture subtle microstructural damage that may underlie persistent symptoms \citep{lee2020advanced}.To address this limitation, diffusion tensor imaging (DTI) provides more sensitive and quantitative information by characterizing white matter microstructural integrity. For example, it can reveal alterations associated with axonal organization, demyelination, and connectivity disruption by mapping water diffusion along white matter tracts. Building upon DTI, more advanced diffusion models, such as diffusion kurtosis imaging (DKI), further probe tissue heterogeneity and microstructural complexity. Compared to DTI, DKI is sensitive to microstructural complexity potentially associated with axonal beading, cytotoxic edema and microglial activation, all of which are common in TBI. In practice, DKI requires higher b-values, a larger number of diffusion-encoding directions, and high image quality to reliably capture non-Gaussian behavior. This poses a challenge for MRI acquisition, especially for clinical settings, where reproducibility and stability are limited. Indeed, \citet{mahan2021evaluation} noted that increases, decreases, and no changes in DTI metrics have been reported in TBI studies. \citet{huang2022white} explained these inconsistencies by highlighting the roles of injury severity, injury type, time since injury, and sample characteristics. Through longitudinal monitoring of early and follow-up examinations, \citet{muftuler2020serial} demonstrated that DKI was more sensitive than DTI and notably revealed delayed microstructural alterations emerging after injury. Although our understanding of the full spectrum of brain tissue injury and recovery processes remains limited, there is a general consensus that higher spatial resolution and signal-to-noise ratio (SNR) improve the ability to delineate fine-scale white matter architecture and reduce partial volume effects \citep{kornaropoulos2022sensitivity}. \citet{choi2011dti} demonstrated that 7T provides a substantially higher SNR, resulting in more stable and less variable quantitative measurements. This technical advantage creates the opportunity to detect more subtle microstructural abnormalities. Conventional strategies for analyzing DTI, such as GLM-based voxel-wise analyses and tract-based spatial statistics, are primarily designed for group studies \citet{hashim2017investigating,gencc2025early}. Their focus on population-level inference and sensitivity to cohort heterogeneity make them less suited to detecting subtle abnormalities at the individual level.

To overcome these limitations, \citet{attye2021tractlearn} proposed TractLearn, which learns tract-specific latent representations from healthy controls and back-projects subject-level reconstruction deviations onto diffusion-parametric images. Building on this approach, \citet{roger2022leveraging} applied TractLearn in patients with chronic temporal lobe epilepsy, and \citet{clemente2023individualised} used TractLearn to profile moderate to severe TBI, demonstrating that TractLearn can generate individualized white matter signatures and reveal inter-patient heterogeneity that remains hidden in conventional group-based analyses. However, the original TractLearn framework models individual deviations relative to healthy variability but does not explicitly incorporate patient-group information or optimize patient–control separability. As a result, although TractLearn may be sensitive to individual deviations from the healthy manifold, these deviations are not necessarily specific to patient-related microstructural alterations. Taken together, these studies reveal a key gap: although advanced diffusion models and 7T MRI improve sensitivity to microstructural damage, current individualized analytical frameworks remain challenged by the balance between sensitivity to subject-specific deviations and specificity for patient-related alterations, particularly in studies with limited sample sizes. This raises the question of whether detected abnormalities reflect true pathology or normal variation. Addressing this limitation is essential for translating diffusion-MRI findings into clinically meaningful patient-level assessments. In this context, we investigate DKI-derived white-matter alterations in athletes with persistent post-concussive symptoms (PPCS) and athletes exposed to repetitive head impacts (RHI), using state-of-the-art 7T DKI. To improve the balance between sensitivity to individualized abnormalities and specificity for patient-related alterations, we introduce TractSpLearn, an extension of TractLearn that incorporates both between-group differences and within-group variability into the learning framework. We compared TractSpLearn with the original TractLearn across multiple white-matter tracts and diffusion-derived parameters to evaluate whether integrating individual deviations and patient-group information improves the detection of patient-related microstructural alterations in limited-sample settings.


\section{Materials and Methods}
\subsection{Participants}
This retrospective study evaluated the performance of TractLearn and TractSpLearn on a cohort of fifty-six participants. Ethical approval was obtained from the institutional review board of Lund University, Sweden (Dnr 2017/1049), and all procedures complied with the Declaration of Helsinki. Written informed consent was obtained from all participants. The cohort consisted of 20 healthy controls (HC), 16 athletes with persistent post-concussive syndromes (PPCS), and 20 athletes with a history of repeated head injuries (RHI). 

The HC group comprised individuals without a history of competitive contact sports who were age- and sex-matched to the athletes, each engaging in $\geq$ 3 exercise sessions per week for the preceding five years. The PPCS group included former elite athletes who continued to experience post-concussive symptoms for at least six months after concussion, with a severity sufficient to prevent return to sport and to limit work, school, or regular social activities. The RHI group consisted of active athletes with a documented history of recurrent head injuries and ongoing exposure to repetitive head impacts during sport. 

\subsection{MRI acquisition}
Both structural and diffusion-weighted imaging (DWI) were acquired on a 7T Philips Achieva scanner. High-resolution anatomical images were obtained using a 3D T1-weighted magnetization-prepared rapid gradient-echo (MPRAGE) sequence with the following parameters: field of view (FOV): 230 × 230 × 180 mm\(^3\), resolution: 0.80 × 0.80  × 0.80 mm\(^3\), repetition time (TR): 8.00 ms, echo time (TE): 1.97 ms. DWI data were collected with FOV = 224 × 224 × 110 mm\(^3\), resolution = 2 × 2 × 2 mm\(^3\), TR = 9200 ms, TE = 65 ms, using multiple diffusion-encoding directions: two b = 0 s/mm\(^2\) volumes with reversed phase-encoding directions for distortion correction, 6 directions at b = 100 s/mm\(^2\), 6 directions at b = 500 s/mm\(^2\), 10 directions at b = 1000 s/mm\(^2\), and 30 directions at b = 2000 s/mm\(^2\). 

\subsection{DWI preprocessing and DKI fitting}
DWI data were preprocessed using a standardized pipeline to ensure data quality and consistency across participants \citep{gard2024widespread}. The pipeline included: (i) denoising using Marchenko--Pastur principal component analysis \citep{veraart2016denoising}; (ii) Gibbs ringing correction using a non-iterative approach \citep{kellner2016gibbs}; (iii) TOPUP susceptibility distortion correction using a reversed diffusion enconding b=0 acquisition, and (iv) eddy current and motion correction. After preprocessing, diffusion and kurtosis metrics, including fractional anisotropy (FA), mean diffusivity (MD), axial diffusivity (AD), radial diffusivity (RD), mean kurtosis (MK), axial kurtosis (AK), and radial kurtosis (RK), were estimated using weighted linear least squares as implemented in DIPY (DiffusionKurtosisModel) \citep{garyfallidis2014dipy}.

\subsection{TractSplearn : TractLearn based from work with individual and group criterion}
The proposed method, referred to as TractSpLearn, builds upon the original TractLearn framework developed by \citep{attye2021tractlearn}. TractLearn and TractSpLearn were applied independently to each white-matter tract and each diffusion-derived parameter. White-matter tracts were segmented using TractSeg resulting in 72 tract-specific regions for each DKI parameter\citep{Wasserthal2018TractSeg} .

The patient (P) cohorts,  PPCS and RHI, were analyzed separately using the same HC cohort as the reference group. For each cohort, we implemented a paired leave-one-out (LOO) evaluation strategy, as illustrated in figure 1. In each fold, one HC--PAT pair, matched by subject index, was reserved for testing. For PPCS, 16 HC--patient pairs were evaluated, with the remaining four HCs retained in each training fold. For RHI, all 20 HC--patient pairs were evaluated. The same held-out pair was evaluated using both TractLearn and TractSpLearn. TractLearn was trained using only the remaining HC, whereas TractSpLearn was trained using the remaining HC together with the remaining patients. 
\begin{figure}
    \centering
    \includegraphics[width=0.9\linewidth]{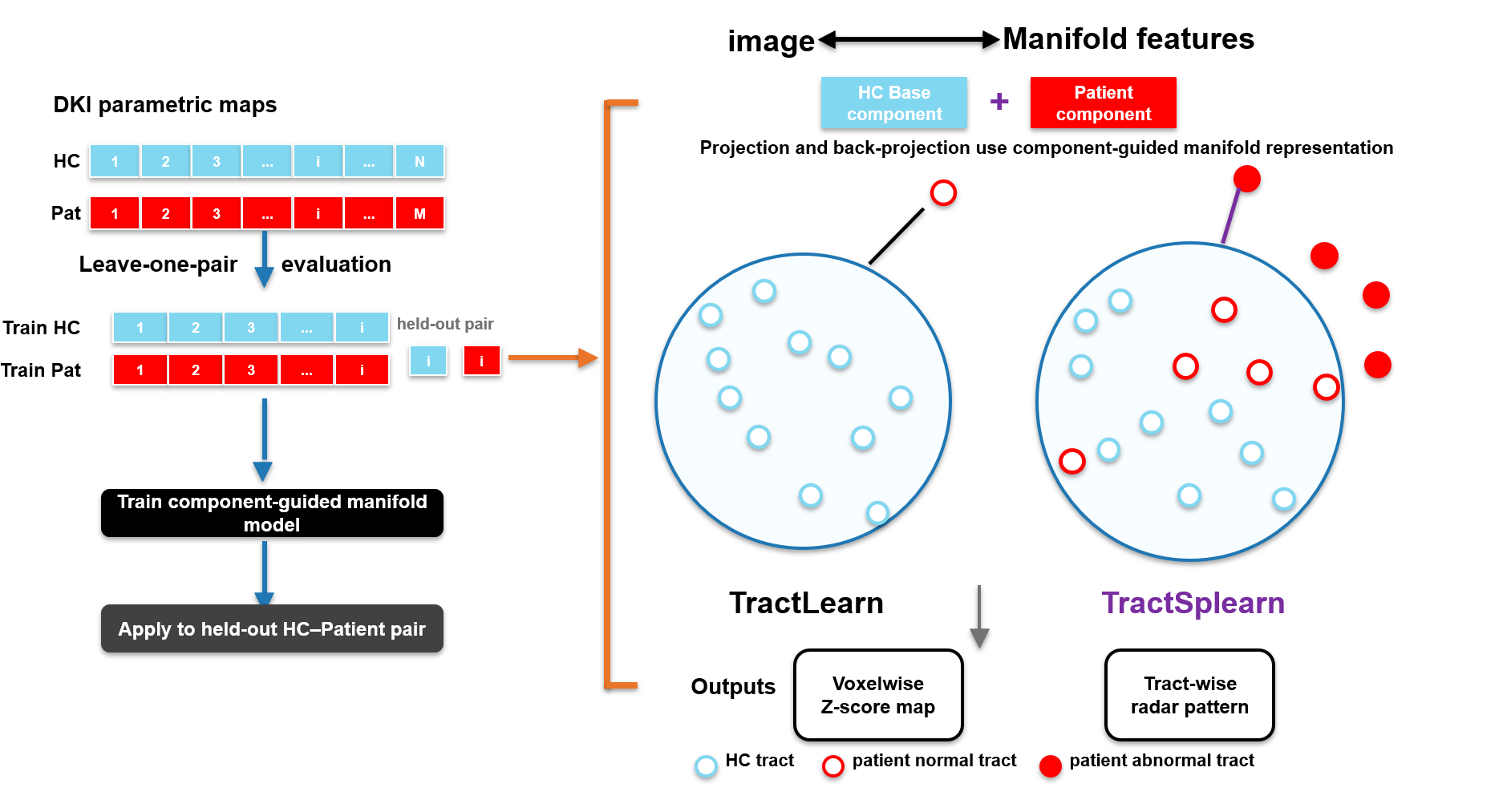}
    \caption{Flowchart of TractLearn and of TractSpLearn and key differences between TractLearn and TractSpLearn}
    \label{fig:placeholder}
\end{figure}
For each LOO training fold, TractLearn  and TractSplearn were applied independently to each white-matter tract and each diffusion-derived parameter. The processing pipeline consisted of manifold feature extraction, forward mapping, back-projection, and healthy-reference z-score calculation.
\paragraph{Features extraction}
Unlike the original TractLearn framework, in which the manifold features were derived exclusively from HCs, TractSpLearn extracted shared manifold features from both HCs and patients in the training set. For each subject \(i\) and white-matter tract \(j\), a set of manifold features \(f_{i,j}\) was extracted from the high-dimensional DKI parameter vector \(p_{i,j}\), where
\begin{equation}
p \in P =
\{
\text{FA}, \text{MD}, \text{AD}, \text{RD}, \text{MK}, \text{AK}, \text{RK}
\}.
\end{equation}
The manifold features were defined as
\begin{equation}
f_{i,j}
=
\mathrm{U}
\left(
p_{i,j}
\right),
\end{equation}
where \(\mathrm{U}(\cdot)\) denotes the UMAP manifold-learning method \citep{mcinnes2018umap}. UMAP projected the HC and patient training data into a shared latent manifold space. The resulting representation captures the common tract structure and inter-subject variability present in both groups.
\paragraph{Forward mapping (projection to the manifold)}
To enable inference on unseen subjects, a forward nonlinear regression model was trained to estimate the manifold coordinates directly from \(p_{i,j}\):
\begin{equation}
\hat{f}_{i,j}
=
\Psi
\left(
p_{i,j}
\right).
\end{equation}
The forward mapping was parameterized as a healthy-anchored base component
together with an additional patient-related component:
\begin{equation}
\Psi
\left(
p_{i,j}
\right)
=
\Psi_{\mathrm{base}}
\left(
p_{i,j}
\right)
+
y_i
\Psi_{\mathrm{component}}
\left(
p_{i,j}
\right)
+
\epsilon_{i,j},
\end{equation}
where \(y_i \in \{0,1\}\) indicates the group label, with \(y_i=0\) for HCs
and \(y_i=1\) for patients. Here, \(\Psi_{\mathrm{base}}(\cdot)\) represents the healthy-anchored base
component, \(\Psi_{\mathrm{component}}(\cdot)\) represents the additional patient-related component, and \(\epsilon_{i,j}\) is the residual term.
For each held-out subject, two latent representations were generated independently of the subject's group label. The forced HC representation was defined as
\begin{equation}
\hat{f}^{\mathrm{HC}}_{i,j}
=
\Psi_{\mathrm{base}}
\left(
p_{i,j}
\right),
\end{equation}
whereas the forced patient representation was defined as

\begin{equation}
\hat{f}^{\mathrm{PAT}}_{i,j}
=
\Psi_{\mathrm{base}}
\left(
p_{i,j}
\right)
+
\Psi_{\mathrm{component}}
\left(
p_{i,j}
\right).
\end{equation}

\paragraph{Back-projection (reconstruction in parameter space)}
The estimated HC and patient manifold representations were reconstructed back into the original DKI parameter space using a shared inverse nonlinear regression model \(\Phi(\cdot)\):
\begin{equation}
\hat{p}^{\mathrm{HC}}_{i,j}
=
\Phi
\left(
\hat{f}^{\mathrm{HC}}_{i,j}
\right),
\end{equation}
and
\begin{equation}
\hat{p}^{\mathrm{PAT}}_{i,j}
=
\Phi
\left(
\hat{f}^{\mathrm{PAT}}_{i,j}
\right).
\end{equation}
Therefore, the difference between the two reconstructed tract vectors originated from their latent representations rather than from separate group-specific back-projection models. The corresponding reconstruction residuals were defined as
\begin{equation}
r^{\mathrm{HC}}_{i,j}
=
p_{i,j}
-
\hat{p}^{\mathrm{HC}}_{i,j},
\end{equation}
and
\begin{equation}
r^{\mathrm{PAT}}_{i,j}
=
p_{i,j}
-
\hat{p}^{\mathrm{PAT}}_{i,j}.
\end{equation}
\paragraph{Z-score calculation}
Rather than directly subtracting the two reconstructed tract vectors, TractSpLearn compared the absolute reconstruction residuals obtained from the HC and patient branches. For subject \(i\), tract \(j\), and voxel \(x\), the signed reconstruction-preference evidence was calculated as
\begin{equation}
S_{i,j}(x)
=
\left|
r^{\mathrm{HC}}_{i,j}(x)
\right|
-
\left|
r^{\mathrm{PAT}}_{i,j}(x)
\right|.
\end{equation}
Equivalently,
\begin{equation}
S_{i,j}(x)
=
\left|
p_{i,j}(x)
-
\hat{p}^{\mathrm{HC}}_{i,j}(x)
\right|
-
\left|
p_{i,j}(x)
-
\hat{p}^{\mathrm{PAT}}_{i,j}(x)
\right|.
\end{equation}
A positive value indicates that the patient branch produced a smaller reconstruction error than the HC branch, whereas a negative value indicates that the HC branch provided the better reconstruction. A voxel-wise normative reference was estimated from the signed evidence maps of the training HC using a leave-one-out procedure. The mean and standard deviation of the HC leave-one-out evidence were calculated as
\begin{equation}
\mu_{j,p}(x)
=
\mathrm{Mean}
\left(
S^{\mathrm{LOO}}_{h,j}(x)
\right),
\qquad
h \in \mathrm{HC}_{\mathrm{train}},
\end{equation}
and
\begin{equation}
\sigma_{j,p}(x)
=
\mathrm{SD}
\left(
S^{\mathrm{LOO}}_{h,j}(x)
\right),
\qquad
h \in \mathrm{HC}_{\mathrm{train}}.
\end{equation}
For each held-out subject, the corresponding TractSpLearn Z-score was calculated as
\begin{equation}
Z_{i,j}(x)
=
\frac{
S_{i,j}(x)
-
\mu_{j,p}(x)
}{
\sigma_{j,p}(x)
},
\qquad
i \in \mathrm{test}.
\end{equation}
\subsection{Evaluation metrics}
As described in Fig. 1, TractLearn and TractSpLearn were evaluated at group and individual levels. The global analysis quantified the overall abnormality burden for each diffusion parameters (and further for each tracts). The individual-level analysis characterized subject-specific abnormality patterns.
\paragraph{Group level}
For each held-out subject, voxel-wise Z-score maps were obtained independently for each white-matter tract and diffusion parameter. Tracts containing fewer than 50 valid voxels in the original tract support were excluded from the quantitative analysis. In addition, voxels with extreme standardized values ($|Z|>10$) were excluded before quantitative averaging to reduce the influence of unstable Z-scores potentially arising from very small normative variance. The tract-size criterion was determined before this voxel-level exclusion.
For subject $i$, tract $j$, diffusion parameter $p\in\mathcal{P}$, and voxel $x\in V_{i,j}^{*}$, where $V_{i,j}^{*}$ denotes the retained voxels after quality-control filtering, the continuous abnormality evidence was defined as
\begin{equation}
S_{i,j,p}(x)=
\begin{cases}
|Z_{i,j,p}(x)|,
& \text{TractLearn}, \\[4pt]
\max\left(Z_{i,j,p}(x),0\right),
& \text{TractSpLearn}.
\end{cases}
\label{eq:continuous_evidence}
\end{equation}
For TractLearn, both positive and negative deviations from the HC normative reference contributed to the abnormality evidence, whereas for TractSpLearn,only positive deviations indicating increased preference for the patient-related reconstruction were retained.
For each subject and diffusion parameter, the global abnormality evidence was calculated by averaging the continuous evidence across all retained voxels from all eligible white-matter tracts. Thus, the global measure was voxel-weighted, with each retained voxel contributing equally to the final score. HC and patient distributions were compared independently for the PPCS and RHI cohorts using two-sided Mann--Whitney U tests with Benjamini–Hochberg false discovery rate for each DKI parameter. Individual subject values together with group means and standard deviations were reported.
\paragraph{Individual-level abnormality patterns}
To characterize subject-specific white-matter abnormalities, voxel-wise Z-score maps were converted into binary abnormality maps using method-specific criteria:
\begin{equation}
B_{i,j,p}(x)=
\begin{cases}
\mathbb{I}\left(|Z_{i,j,p}(x)|>T\right),
& \text{TractLearn}, \\[4pt]
\mathbb{I}\left(Z_{i,j,p}(x)>T\right),
& \text{TractSpLearn},
\end{cases}
\end{equation}
where $T=4.25$ denotes the predefined voxel-wise Z-score threshold. For TractLearn, both positive and negative deviations from the healthy reference were considered abnormal, whereas for TractSpLearn only positive Z-scores were retained because positive values represent increased preference for the patient-related representation relative to the healthy reference. For each subject and diffusion parameter, a tract-wise abnormality percentage was calculated as the proportion of supra-threshold voxels among all retained voxels within each white-matter tract. This percentage was used to summarize the extent of subject-specific tract involvement. Tracts containing fewer than 50 retained voxels were excluded from the tract-level analysis to reduce instability associated with very small tract representations. 
\section{Result}
\subsection{Group-level comparison of abnormality scores}
To compare the abnormality patterns identified by TractLearn and TractSpLearn, we first summarized the subject-level abnormality evidence at the group level. Of the 72 tracts,  29  containing fewer than 50 valid voxels were excluded from subsequent analyses. For each diffusion parameter, voxel-wise evidence was aggregated across the analyzed white matter tracts to obtain a global abnormality score for each subject. Figure 2 shows the group-level distributions for FA, MD, AD, RD, MK, AK, and RK, with the corresponding statistical significance indicated in the figure. 
\begin{figure}
    \centering
    \includegraphics[width=0.9\linewidth]{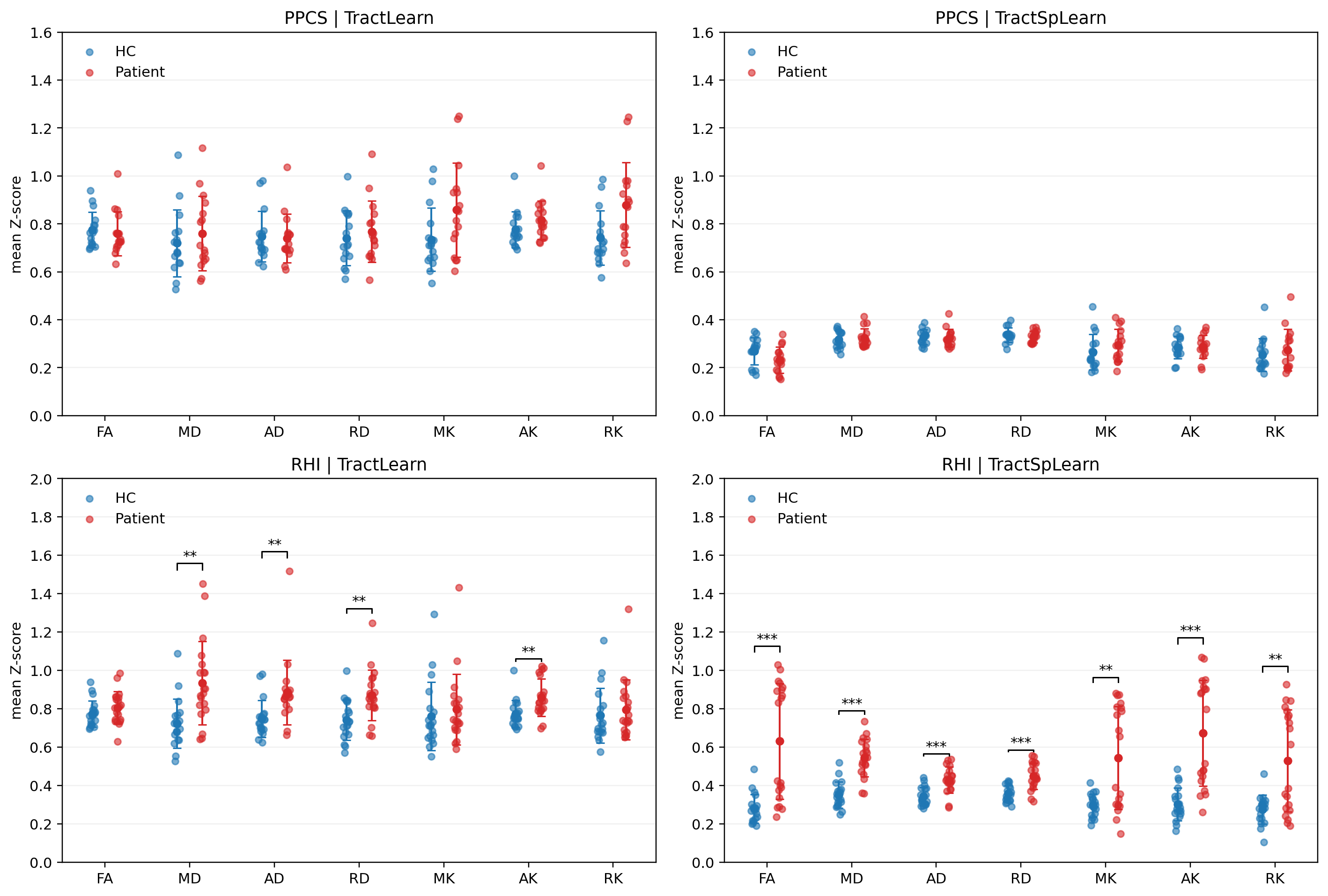}
    \caption{Group-averaged Z-score profiles for HC and patient groups}
\end{figure}
Overall, the pattern of within-cohort HC–patient differences varied between the PPCS and RHI cohorts. In the PPCS cohort, substantial overlap between HC and patients was observed for most diffusion parameters with both methods, indicating relatively weak group-level differences. In contrast, better delineated group differences were observed in the RHI cohort. TractLearn showed significant differences for MD, AD, RD, and AK, whereas TractSpLearn showed significant HC–RHI differences across all seven diffusion parameters, including FA, MD, AD, RD, MK, AK, and RK. Based on the group-level findings, MD and AK were selected for subsequent individual-level visualization and analysis. MD was chosen as a representative conventional diffusion metric, whereas AK was chosen as a representative kurtosis diffusion coefficient showing particularly prominent abnormalities and group separation. In the following analyses, we therefore focused on these two parameters to illustrate subject-specific abnormality patterns and spatial distributions. 
\begin{figure}[htbp]
    \centering
    \includegraphics[width=0.6\linewidth]{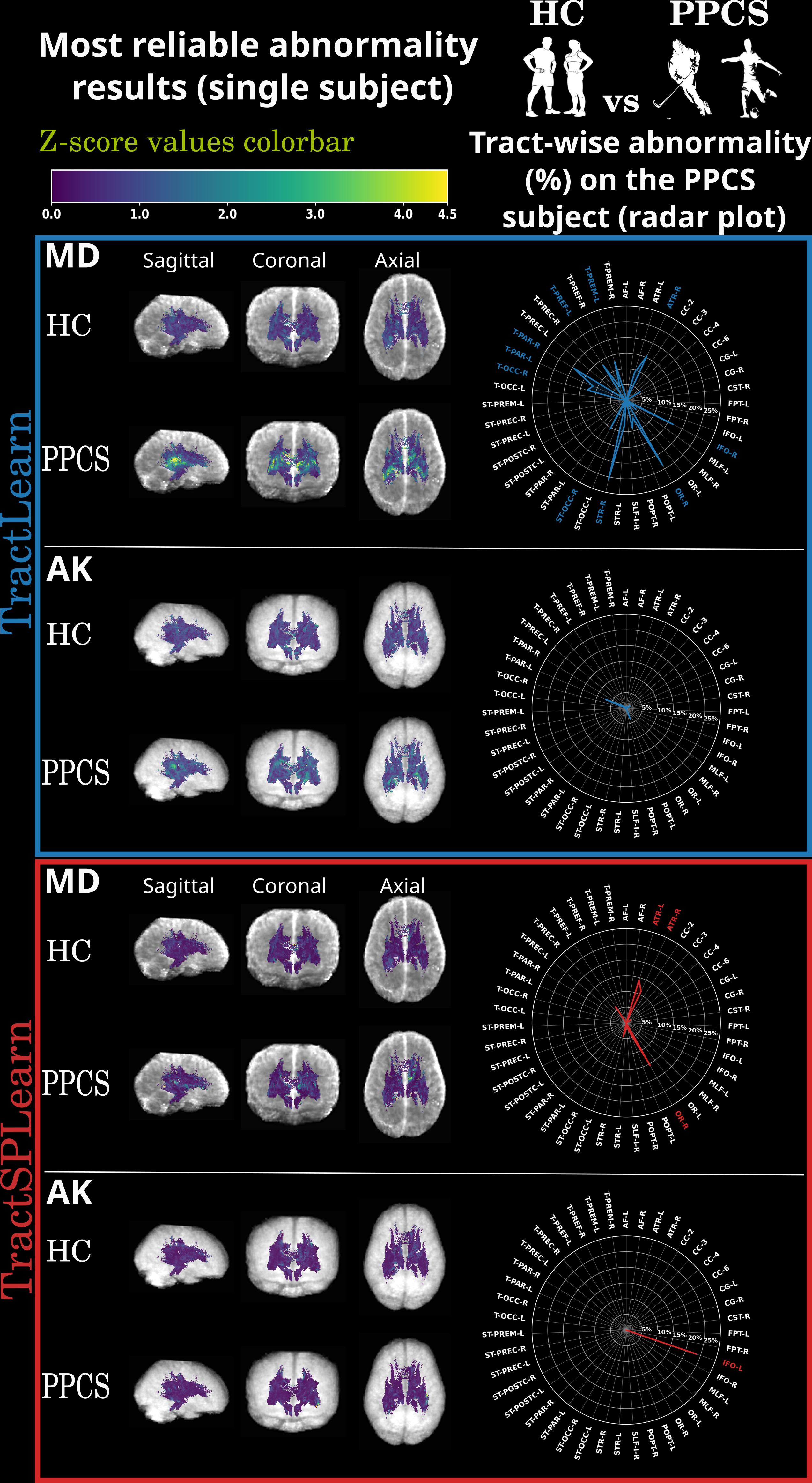}
    \caption{Individual-level tract-wise abnormality patterns in representative
    PPCS subjects. The left blue-framed block shows TractLearn results, and
    the right red-framed block shows TractSpLearn results. MD is shown in the
    upper row and AK in the lower row. For each parameter, sagittal, coronal
    and axial views of the spatial abnormality maps are followed by the
    corresponding tract-wise abnormality profile in a radar plot.}
    \label{fig:individual_PPCS}
\end{figure}

\begin{figure}[htbp]
    \centering
    \includegraphics[width=0.6\linewidth]{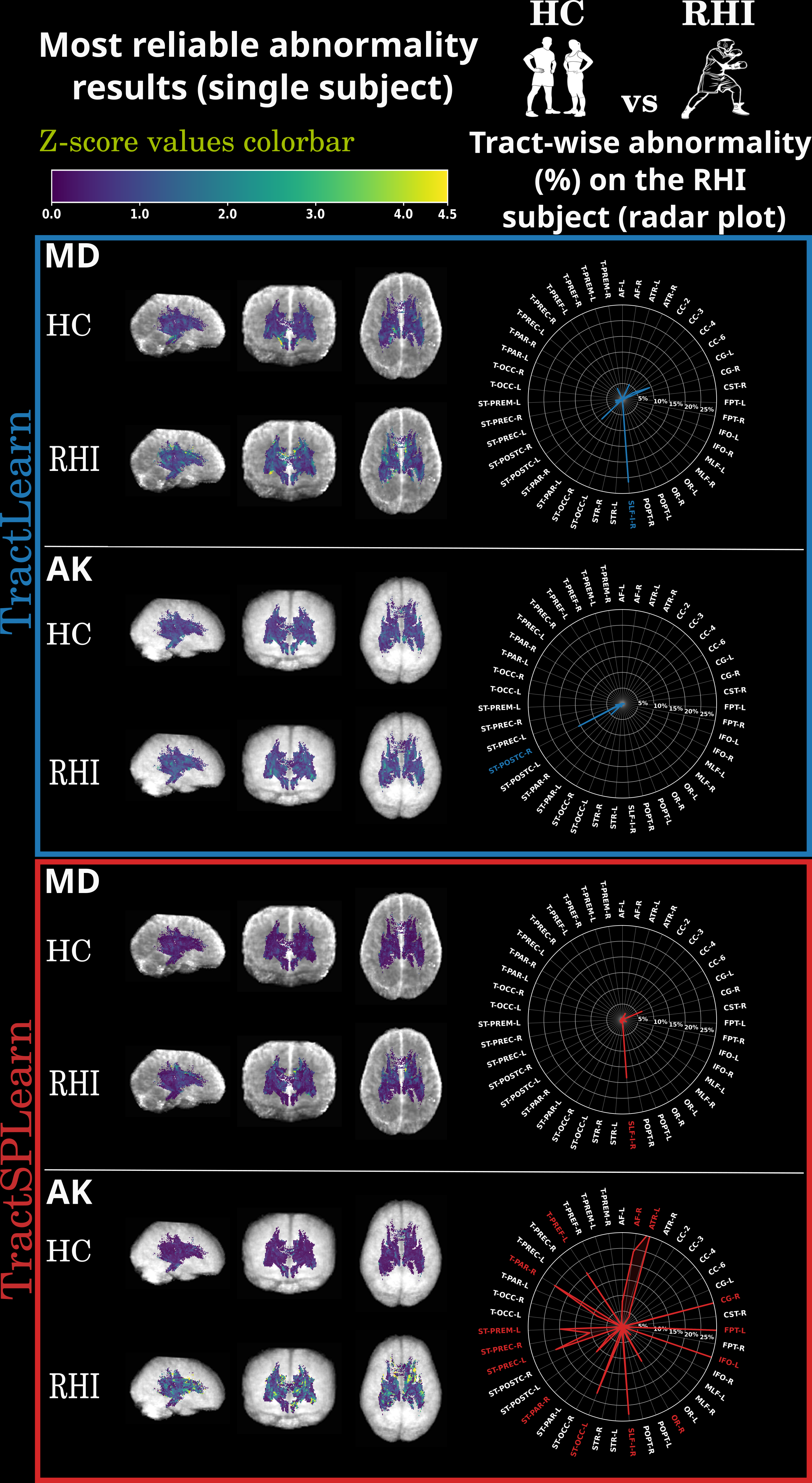}   
    \caption{Individual-level tract-wise abnormality patterns in representative
    RHI subjects. The left blue-framed block shows TractLearn results, and
    the right red-framed block shows TractSpLearn results. MD is shown in the
    upper row and AK in the lower row. For each parameter, sagittal, coronal
    and axial views of the spatial abnormality maps are followed by the
    corresponding tract-wise abnormality profile in a radar plot.}
    \label{fig:individual_RHI}
\end{figure}
Figure 3 and Figure 4 show representative individual results for the PPCS and RHI cohorts. At the individual level, patients generally exhibited more extensive and spatially clustered abnormality patterns than healthy controls, although the magnitude and distribution of these alterations varied across subjects and diffusion parameters. TractLearn also produced visible supra-threshold abnormalities in healthy controls, whereas TractSpLearn generally yielded sparser patient-related evidence in controls while retaining localized abnormalities in patients. This contrast was particularly apparent in the RHI data, where TractSpLearn showed more spatially concentrated patient-related alterations with comparatively limited background evidence in healthy controls. For each representative patient and healthy control, the corresponding radar plots summarize the proportion of supra-threshold voxels within each white matter tract, providing an individual-level profile of tract involvement.

To further evaluate lesion detection at the subject level, we summarized the number of participants with at least one affected white-matter tract and the number of affected tracts per subject across the selected diffusion parameters. A tract was considered affected when at least 15\% of its voxels showed abnormal values according to the predefined Z-score criterion. For the combined analysis across diffusion parameters, the same tract was counted only once when it was identified as abnormal in more than one parameter. Subjects with at least one affected tract were classified as lesion-positive.

\begin{table*}[t]
\centering
\footnotesize
\setlength{\tabcolsep}{2pt}
\renewcommand{\arraystretch}{0.9}
\caption{Individual patient tract results and cohort summary for the PPCS cohort. Here, n denotes the number of abnormal tracts; tract percentages denote the percentage of voxels with abnormal values.}
\label{tab:ppcs-tract-results}
\begin{tabular}{p{1.2cm}p{0.75cm}cp{2.8cm}p{1.0cm}p{0.8cm}@{\hspace{1.2em}}p{0.75cm}cp{2.8cm}p{1.0cm}p{0.8cm}}
\hline
& \multicolumn{5}{c}{\textbf{TractLearn}} & \multicolumn{5}{c}{\textbf{TractSpLearn}} \\
\cline{2-6} \cline{7-11}
Subject & n & Rank & Tract & \shortstack{Tract\\(\%)} & Param. & n & Rank & Tract & \shortstack{Tract\\(\%)} & Param. \\
\hline
PPCS1 & 4 & 1 & OR\_right & 28.333 & MD & 2 & 1 & ST\_PREC\_left & 30.872 & AK \\
 &  & 2 & IFO\_right & 22.973 & MD &  & 2 & STR\_left & 25.843 & AK \\
 &  & 3 & ST\_OCC\_right & 22.807 & MD &  &  &  &  & \\
PPCS2 & -- & -- & -- & -- & -- & 2 & 1 & STR\_left & 23.750 & AK \\
 &  &  &  &  &  &  & 2 & ST\_PREC\_left & 18.750 & AK \\
PPCS3 & -- & -- & -- & -- & -- & -- & -- & -- & -- & -- \\
PPCS4 & -- & -- & -- & -- & -- & -- & -- & -- & -- & -- \\
PPCS5 & -- & -- & -- & -- & -- & -- & -- & -- & -- & -- \\
PPCS6 & -- & -- & -- & -- & -- & 1 & 1 & T\_OCC\_left & 20.548 & AK \\
PPCS7 & -- & -- & -- & -- & -- & -- & -- & -- & -- & -- \\
PPCS8 & 5 & 1 & STR\_right & 25.806 & MD & 2 & 1 & IFO\_left & 23.529 & AK \\
 &  & 2 & OR\_right & 23.810 & MD &  & 2 & OR\_right & 15.556 & MD \\
 &  & 3 & T\_PAR\_right & 19.497 & MD &  &  &  &  & \\
PPCS9 & -- & -- & -- & -- & -- & 2 & 1 & IFO\_left & 31.132 & AK \\
 &  &  &  &  &  &  & 2 & ST\_OCC\_left & 21.978 & AK \\
PPCS10 & -- & -- & -- & -- & -- & 2 & 1 & IFO\_left & 29.508 & AK \\
 &  &  &  &  &  &  & 2 & ST\_OCC\_left & 19.797 & AK \\
PPCS11 & -- & -- & -- & -- & -- & 1 & 1 & ST\_POSTC\_right & 19.380 & AK \\
PPCS12 & -- & -- & -- & -- & -- & 1 & 1 & IFO\_left & 36.441 & AK \\
PPCS13 & -- & -- & -- & -- & -- & -- & -- & -- & -- & -- \\
PPCS14 & -- & -- & -- & -- & -- & -- & -- & -- & -- & -- \\
PPCS15 & -- & -- & -- & -- & -- & 1 & 1 & STR\_left & 27.473 & AK \\
PPCS16 & -- & -- & -- & -- & -- & -- & -- & -- & -- & -- \\
\hline
\multicolumn{1}{l}{PAT summary} & \multicolumn{5}{p{5.7cm}}{\parbox{5.5cm}{\raggedright Positive PATs: 2/16 (12.5\%)\\Mean affected tracts per positive PAT: 4.50 $\pm$ 0.71}} & \multicolumn{5}{p{5.7cm}}{\parbox{5.5cm}{\raggedright \textbf{Positive PATs: 9/16 (56.3\%)\\}Mean affected tracts per positive PAT: 1.56 $\pm$ 0.53}} \\
\hline
\multicolumn{1}{l}{HC summary} & \multicolumn{5}{p{5.7cm}}{\parbox{5.5cm}{\raggedright Positive HCs: 3/16 (18.8\%)\\Mean affected tracts per positive HC: 2.33 $\pm$ 0.58}} & \multicolumn{5}{p{5.7cm}}{\parbox{5.5cm}{\raggedright Positive HCs: 3/16 (18.8\%)\\Mean affected tracts per positive HC: 1.67 $\pm$ 1.15}} \\
\hline
\end{tabular}
\end{table*}

\begin{table*}[p]
\centering
\footnotesize
\setlength{\tabcolsep}{2pt}
\renewcommand{\arraystretch}{0.9}
\caption{Individual patient tract results and cohort summary for the RHI cohort. Here, n denotes the number of abnormal tracts; tract percentages denote the percentage of voxels with abnormal values.}
\label{tab:rhi-tract-results}
\begin{tabular}{p{1.2cm}p{0.75cm}cp{2.8cm}p{1.0cm}p{0.8cm}@{\hspace{1.2em}}p{0.75cm}cp{2.8cm}p{1.0cm}p{0.8cm}}
\hline
& \multicolumn{5}{c}{\textbf{TractLearn}} & \multicolumn{5}{c}{\textbf{TractSpLearn}} \\
\cline{2-6} \cline{7-11}
Subject & n & Rank & Tract & \shortstack{Tract\\(\%)} & Param. & n & Rank & Tract & \shortstack{Tract\\(\%)} & Param. \\
\hline
RHI1 & 1 & 1 & CC\_6 & 18.027 & MD & 3 & 1 & CC\_6 & 30.612 & MD \\
 &  &  &  &  &  &  & 2 & T\_PAR\_right & 19.628 & AK \\
 &  &  &  &  &  &  & 3 & FPT\_left & 16.597 & MD \\
RHI2 & 1 & 1 & CC\_6 & 16.667 & MD & 8 & 1 & ST\_PAR\_left & 34.444 & AK \\
 &  &  &  &  &  &  & 2 & T\_PREC\_left & 32.277 & AK \\
 &  &  &  &  &  &  & 3 & T\_OCC\_right & 31.579 & AK \\
RHI3 & -- & -- & -- & -- & -- & 2 & 1 & T\_PREC\_left & 24.051 & AK \\
 &  &  &  &  &  &  & 2 & FPT\_right & 19.057 & AK \\
RHI4 & -- & -- & -- & -- & -- & 3 & 1 & FPT\_right & 26.304 & AK \\
 &  &  &  &  &  &  & 2 & STR\_left & 19.780 & AK \\
 &  &  &  &  &  &  & 3 & T\_OCC\_left & 15.278 & AK \\
RHI5 & -- & -- & -- & -- & -- & 5 & 1 & ST\_POSTC\_right & 56.044 & AK \\
 &  &  &  &  &  &  & 2 & CC\_2 & 30.000 & AK \\
 &  &  &  &  &  &  & 3 & ST\_PAR\_left & 28.019 & AK \\
RHI6 & -- & -- & -- & -- & -- & 14 & 1 & CG\_right & 40.367 & AK \\
 &  &  &  &  &  &  & 2 & FPT\_left & 33.264 & AK \\
 &  &  &  &  &  &  & 3 & T\_OCC\_left & 29.688 & AK \\
RHI7 & -- & -- & -- & -- & -- & 9 & 1 & CC\_3 & 46.000 & AK \\
 &  &  &  &  &  &  & 2 & T\_PREC\_left & 43.910 & AK \\
 &  &  &  &  &  &  & 3 & ST\_POSTC\_left & 43.243 & AK \\
RHI8 & -- & -- & -- & -- & -- & 1 & 1 & T\_PAR\_right & 21.250 & AK \\
RHI9 & -- & -- & -- & -- & -- & -- & -- & -- & -- & -- \\
RHI10 & -- & -- & -- & -- & -- & 2 & 1 & SLF\_I\_right & 29.825 & AK \\
 &  &  &  &  &  &  & 2 & POPT\_left & 16.166 & AK \\
RHI11 & -- & -- & -- & -- & -- & 11 & 1 & ST\_POSTC\_left & 40.000 & AK \\
 &  &  &  &  &  &  & 2 & ST\_PAR\_left & 39.950 & AK \\
 &  &  &  &  &  &  & 3 & CC\_3 & 37.037 & AK \\
RHI12 & 6 & 1 & OR\_left & 41.791 & MD & 3 & 1 & T\_PREC\_left & 35.577 & AK \\
 &  & 2 & OR\_right & 34.921 & MD &  & 2 & FPT\_right & 25.558 & AK \\
 &  & 3 & ST\_OCC\_left & 34.682 & MD &  & 3 & POPT\_left & 25.467 & AK \\
RHI13 & -- & -- & -- & -- & -- & 11 & 1 & AF\_left & 49.375 & AK \\
 &  &  &  &  &  &  & 2 & CC\_6 & 41.228 & AK \\
 &  &  &  &  &  &  & 3 & T\_PREC\_left & 38.722 & AK \\
RHI14 & -- & -- & -- & -- & -- & -- & -- & -- & -- & -- \\
RHI15 & 2 & 1 & SLF\_I\_right & 28.302 & MD & 8 & 1 & STR\_left & 48.837 & AK \\
 &  & 2 & ST\_POSTC\_right & 15.504 & AK &  & 2 & FPT\_left & 35.204 & AK \\
 &  &  &  &  &  &  & 3 & IFO\_left & 33.051 & AK \\
RHI16 & -- & -- & -- & -- & -- & 2 & 1 & T\_PREC\_left & 33.234 & AK \\
 &  &  &  &  &  &  & 2 & SLF\_I\_right & 23.214 & AK \\
RHI17 & 1 & 1 & CC\_6 & 17.164 & MD & 13 & 1 & T\_OCC\_right & 35.443 & AK \\
 &  &  &  &  &  &  & 2 & ST\_PAR\_left & 34.422 & AK \\
 &  &  &  &  &  &  & 3 & T\_PREC\_left & 30.973 & AK \\
RHI18 & 1 & 1 & T\_OCC\_left & 17.808 & MD & 5 & 1 & ST\_PREC\_right & 29.762 & AK \\
 &  &  &  &  &  &  & 2 & FPT\_right & 28.352 & AK \\
 &  &  &  &  &  &  & 3 & OR\_left & 27.941 & AK \\
RHI19 & 2 & 1 & SLF\_I\_right & 26.415 & MD & 11 & 1 & IFO\_left & 33.871 & AK \\
 &  & 2 & ST\_POSTC\_right & 15.625 & AK &  & 2 & ATR\_left & 32.143 & AK \\
 &  &  &  &  &  &  & 3 & FPT\_left & 31.990 & AK \\
RHI20 & -- & -- & -- & -- & -- & 8 & 1 & STR\_left & 52.500 & AK \\
 &  &  &  &  &  &  & 2 & STR\_right & 30.851 & AK \\
 &  &  &  &  &  &  & 3 & AF\_right & 29.655 & AK \\
\hline
\multicolumn{1}{l}{PAT summary} & \multicolumn{5}{p{5.7cm}}{\parbox{5.5cm}{\raggedright Positive PATs: 7/20 (35.0\%) \\Mean affected tracts per positive PAT: 2.00 $\pm$ 1.83}} & \multicolumn{5}{p{5.7cm}}{\parbox{5.5cm}{\raggedright \textbf{Positive PATs: 18/20 (90.0\%)\\}Mean affected tracts per positive PAT: 6.61 $\pm$ 4.23}} \\
\hline
\multicolumn{1}{l}{HC summary} & \multicolumn{5}{p{5.7cm}}{\parbox{5.5cm}{\raggedright Positive HCs: 3/20 (15.0\%)\\Mean affected tracts per positive HC: 2.33 $\pm$ 0.58}} & \multicolumn{5}{p{5.7cm}}{\parbox{5.5cm}{\raggedright Positive HCs: 11/20 (55.0\%)\\Mean affected tracts per positive HC: 1.27 $\pm$ 0.47}} \\
\hline
\end{tabular}
\end{table*}
As shown in Tables1 and Table 2, TractSpLearn increased patient-level lesion detection in both cohorts. The mean and standard deviation (SD) of affected tract counts were calculated only among lesion-positive subjects.

In the PPCS cohort, patient positivity increased from 2/16 (12.5\%) with TractLearn to 9/16 (56.3\%) with TractSpLearn, while the mean affected tract count among positive patients decreased from $4.50 \pm 0.71$ to $1.56 \pm 0.53$. HC positivity remained unchanged at 3/16 (18.8\%). Among lesion-positive HCs, the mean affected tract count was $2.33 \pm 0.58$ with TractLearn and $1.67 \pm 1.15$ with TractSpLearn.

In the RHI cohort, patient positivity increased from 7/20 (35.0\%) with TractLearn to 18/20 (90.0\%) with TractSpLearn, accompanied by an increase in the mean affected tract count among positive patients from $2.00 \pm 1.83$ to $6.61 \pm 4.23$. HC positivity also increased, from 3/20 (15.0\%) to 11/20 (55.0\%). However, TractSpLearn-positive HCs had relatively few affected tracts, with a mean of $1.27 \pm 0.47$ per positive HC , compared with $2.33 \pm 0.58$ with TractLearn.

Across cohorts, TractSpLearn showed a marked increase in patient-level detection from PPCS to RHI (56.3\% to 90.0\%), alongside an increase in the mean number of affected tracts among positive patients (1.56 to 6.61). In TractLearn, patient-level detection also increased from PPCS to RHI (12.5\% to 35.0\%), while the mean number of affected tracts among positive patients decreased from 4.50 to 2.00.

\subsection{Exploratory analysis of recurrent abnormality patterns }
\begin{figure}[htbp]
    \centering
    \includegraphics[width=0.9\linewidth]{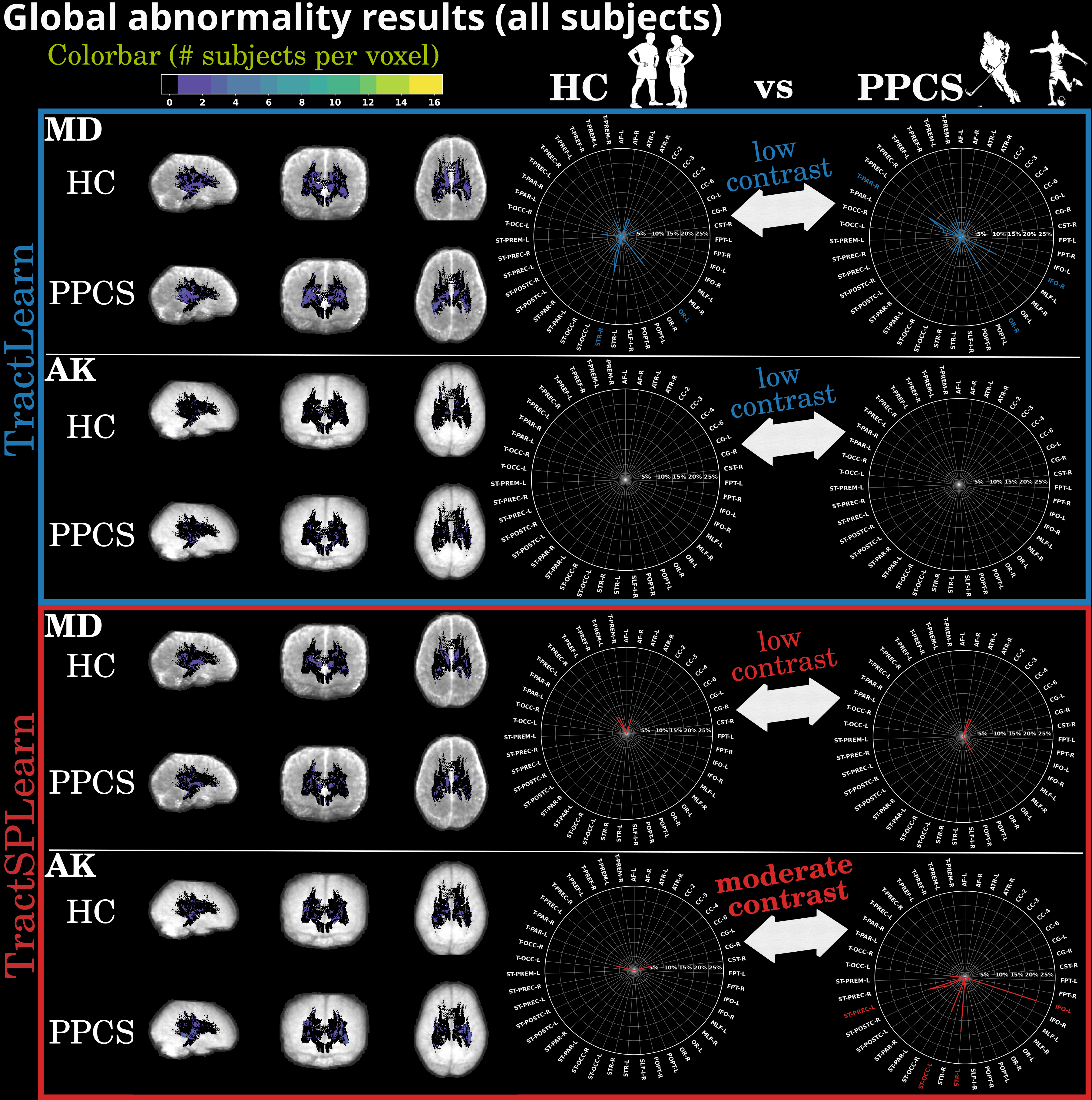}
    \caption{
    Group-level spatial and tract-wise frequency of white matter abnormalities in the PPCS cohort.
    Blue and red outlines indicate TractLearn and TractSpLearn, respectively.
    For each method, the upper and lower blocks show MD and AK; within each block, the upper and lower rows show HCs and PATs.
    Spatial maps show participant counts across sagittal, coronal, and axial views; radar plots show the frequency of abnormality in each tract.
    }
    \label{fig:group_ppcs}
\end{figure}

\begin{figure}[htbp]
    \centering
    \includegraphics[width=0.9\linewidth]{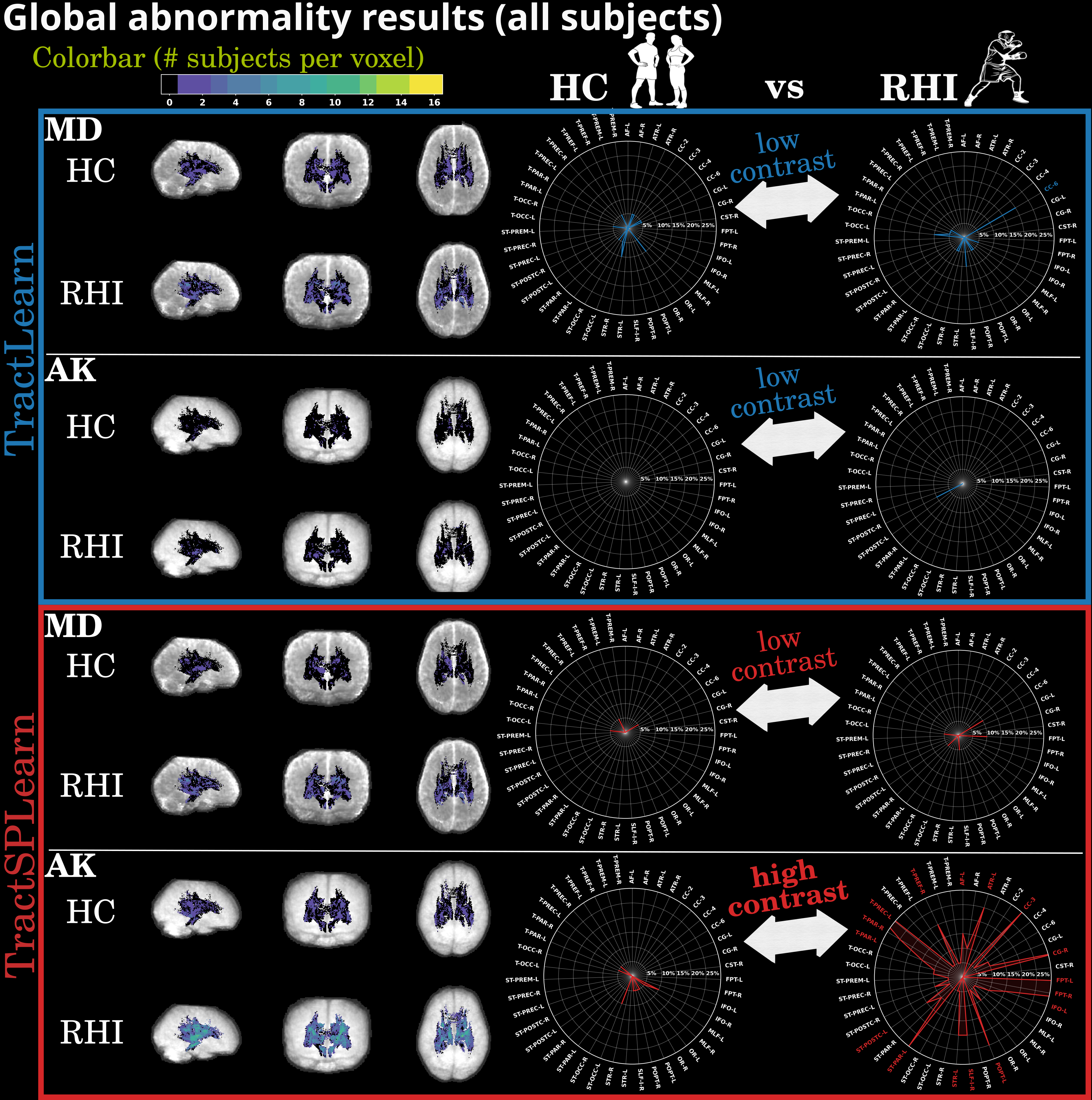}
    \caption{
    Group-level spatial and tract-wise frequency of white matter abnormalities in the RHI cohort.
    Blue and red outlines indicate TractLearn and TractSpLearn, respectively.
    For each method, the upper and lower blocks show MD and AK; within each block, the upper and lower rows show HCs and PATs.
    Spatial maps show participant counts across sagittal, coronal, and axial views; radar plots show the frequency of abnormality in each tract.
    }
    \label{fig:group_rhi}
\end{figure}
To further explore whether TBI was associated with recurrent patterns of white matter abnormalities across individuals, we performed an exploratory group-level analysis of the spatial and tract-wise distributions of detected abnormalities in both the PPCS and RHI cohorts. Figure~5 shows the PPCS results, and Figure~6 shows the RHI results. For each cohort, the figures present spatial abnormality maps in sagittal, coronal, and axial views, together with tract-wise radar plots for the HC and PAT groups. Results are shown for MD and AK using TractLearn and TractSpLearn; the HC results are included as a reference for comparison with the patient group. The spatial frequency maps show the number of participants exhibiting an abnormality at each voxel, while the radar plots summarize the frequency with which each white matter tract was marked as abnormal across participants.

In the PPCS cohort, patient abnormalities were relatively limited. TractLearn abnormalities were observed mainly for MD, whereas TractSpLearn showed abnormalities for both MD and AK. For AK with TractSpLearn, recurrent tract-wise involvement was observed in the left inferior occipito-frontal fascicle (IFO-L), left striato-precentral tract (ST-PREC-L), left striato-occipital tract (ST-OCC-L), and left superior thalamic radiation (STR-L).

In the RHI cohort, the spatial frequency maps showed more extensive recurrent abnormalities than in PPCS. AK abnormalities were broadly distributed across bilateral deep and periventricular white matter, while MD abnormalities also recurred across multiple regions. TractLearn identified more extensive MD abnormalities, whereas TractSpLearn showed recurrent abnormalities in both MD and AK. In the RHI patient group, the TractSpLearn MD radar plot showed frequent involvement across several tracts, including the left anterior thalamic radiation (ATR-L), corpus callosum segment 3 (CC-3), right cingulum (CG-R), bilateral fronto-pontine tracts (FPT-L and FPT-R), left parieto-occipital pontine tract (POPT-L), left superior thalamic radiation (STR-L), right thalamo-prefrontal tract (T-PREF-R), left thalamo-precentral tract (T-PREC-L), right thalamo-parietal tract (T-PAR-R), left striato-parietal tract (ST-PAR-L), and left striato-postcentral tract (ST-POSTC-L). Overall, these exploratory results indicate that recurrent abnormalities differed between cohorts and metrics, with TractLearn showing more extensive MD abnormalities and TractSpLearn identifying abnormalities in both MD and AK.

\section{Discussion}
In this study, we proposed TractSpLearn, a tract-wise framework for individualized characterization of white matter abnormalities by explicitly incorporating patient-related information into the modeling process. Unlike the original TractLearn framework, which identifies individual abnormalities primarily according to deviations from a healthy reference, TractSpLearn models both shared structural characteristics and an additional patient-related component. For an unseen subject, the resulting abnormality measure therefore reflects the relative preference between the healthy and patient representations rather than deviation from the healthy population alone. This formulation provides a more disease-informed characterization of individualized diffusion abnormalities while retaining healthy controls as the normative reference.

The main finding was that HC–patient group differences were modest in PPCS but more clearly delineated in RHI. In RHI, TractLearn showed significant differences for MD, AD, RD, and AK, whereas TractSpLearn showed significant differences for all seven diffusion parameters. At the subject level, TractSpLearn yielded positive findings in more patients in both cohorts; however, HC positivity in RHI also increased from 15.0 \% with TractLearn to 55.0 \% with TractSpLearn. Therefore, the higher proportion of positive patients should not be interpreted as improved specificity, and diagnostic accuracy cannot be inferred without an independent ground truth. A key conceptual difference between TractSpLearn is that the sign of the resulting score carries meaningful information. In TractLearn, the magnitude of the deviation is the primary quantity of interest, and abnormalities can therefore be characterized using the absolute Z-score. In TractSpLearn, however, a positive score indicates that the observed tract profile is better explained by the patient-related representation than expected from the healthy reference distribution. Conversely, a negative score does not represent an abnormality of equal but opposite biological meaning. Therefore, only positive TractSpLearn Z-scores were considered as patient-related abnormality evidence in the present analysis.

This pattern was most evident in RHI, where TractSpLearn revealed significant group differences across a broader range of diffusion measures. Among the diffusion-derived parameters, the abnormalities identified by TractSpLearn were particularly prominent in AK and MD. This finding is consistent with previous diffusion MRI studies showing that both kurtosis- and diffusivity-based measures can be sensitive to microstructural alterations following TBI. For example, longitudinal studies of sports-related concussion have reported significant changes in MD after injury, with some abnormalities persisting beyond clinical symptom resolution \citep{wu2020longitudinal}. Diffusion kurtosis imaging has also demonstrated sensitivity to TBI-related white-matter alterations and may provide complementary information beyond conventional DTI measures \citep{stokum2015longitudinal}. A comparable observation was reported by Chung et al., who found significantly increased AK across widespread white matter in athletes with sports-related concussion, while conventional diffusion measures showed less extensive group differences \citep{chung2022investigating}. Similarly, Muftuler et al. demonstrated longitudinal changes in kurtosis-based measures following sports-related concussion, supporting the sensitivity of DKI-derived parameters to subtle and evolving white-matter alterations after head injury \citep{muftuler2020serial}.

The exploratory spatial and tract-wise analyses showed different patterns across cohorts and methods. In PPCS, recurrent patient abnormalities were relatively limited. TractLearn abnormalities were observed mainly for MD, whereas TractSpLearn showed recurrent abnormalities in both MD and AK. For AK with TractSpLearn, the radar plot highlighted the left inferior fronto-occipital fasciculus (IFO-L), left striato-precentral tract (ST-PREC-L), left striato-occipital tract (ST-OCC-L), and left superior thalamic radiation (STR-L). In RHI, recurrent abnormalities were more extensive than in PPCS. AK abnormalities were broadly distributed across bilateral deep and periventricular white matter, while MD abnormalities also recurred across multiple regions. TractLearn showed more extensive MD abnormalities, whereas TractSpLearn showed recurrent abnormalities in both MD and AK. These maps provide exploratory summaries of cross-subject recurrence and should not be interpreted as inferential evidence that individual tracts are affected at the population level.

The involvement of these pathways is broadly supported by previous diffusion MRI studies of TBI. Yeh et al. reported longitudinal abnormalities involving the fronto-pontine tract, anterior thalamic radiation, corpus callosum, and other long projection pathways following TBI \citep{yeh2022longitudinal}. Palacios et al. also demonstrated diffusion abnormalities involving the cingulum in TBI and showed that alterations in this pathway were related to memory impairment \citep{palacios2011diffusion}. Consistent with this, Wilde et al. reported reduced FA and increased diffusivity in bilateral cingulum bundles after TBI \citep{wilde2010diffusion}. Corpus callosum involvement has likewise been repeatedly observed following TBI, and longitudinal diffusion studies have identified abnormalities in the callosum together with the cingulum and other major association pathways \citep{bendlin2008longitudinal}. More recently, Goubran et al. reported diffusion and kurtosis abnormalities involving the corpus callosum, superior longitudinal fasciculus, thalamic radiation, and cingulum in concussed football players \citep{goubran2023microstructural}. These reports provide context for the white matter pathways highlighted in our exploratory analyses, but they do not independently validate the abnormalities detected in the present study. The observed patterns should therefore be interpreted as model-derived diffusion abnormality evidence rather than as confirmation of specific pathological processes.

The different performance observed between the RHI and PPCS cohorts further illustrates how the benefit of patient-informed modeling may depend on the consistency of disease-related imaging patterns. Although both TractLearn and TractSpLearn generally showed higher mean abnormality evidence in patients than in healthy controls, the relative patterns across the two patient cohorts differed between the methods. With TractLearn, the mean abnormality levels observed in RHI and PPCS were relatively similar, suggesting that a deviation-based framework primarily captures the extent to which an individual differs from the healthy reference, with limited sensitivity to differences in the underlying patient-related pattern. In contrast, TractSpLearn showed a clearer distinction between the two patient cohorts, with substantially stronger patient-related abnormality evidence in RHI than in PPCS. Nevertheless, TractSpLearn identified abnormality evidence in more PPCS patients than TractLearn, although the mean number of affected tracts among abnormality-positive patients decreased. These findings suggest that explicitly incorporating patient-related information can improve the characterization of white-matter abnormalities in both cohorts, while highlighting cohort-specific abnormality patterns when a more consistent disease-related imaging signature is present.

The more pronounced benefit observed in RHI may be related to differences in the underlying pathological and imaging heterogeneity of the two cohorts. Persistent post-concussive symptoms can arise from heterogeneous pathophysiological mechanisms, and both the location and magnitude of diffusion abnormalities may vary considerably across individuals \citep{bouix2013increased}. Such variability can reduce the consistency of the patient-related pattern available for model learning, particularly in a relatively small cohort, and may explain why the improvement in PPCS was more modest and why substantial overlap with healthy controls remained. Previous diffusion MRI studies of repetitive head impacts have demonstrated white-matter microstructural alterations associated with cumulative head-impact exposure \citep{bazarian2014persistent,brett2021association}. One possible explanation for the greater benefit observed in RHI is that more spatially extensive abnormalities increase the likelihood that some white matter regions are affected across multiple patients. These shared regional features could facilitate learning of the patient-related component in TractSpLearn, even when the overall abnormality patterns differ between individuals. However, this explanation remains hypothetical, as differences in the spatial overlap of abnormalities between RHI and PPCS were not directly assessed in the present study.

The individual spatial maps provide further support this interpretation by demonstrating substantial inter-subject variability in both healthy controls and patients. Even in RHI, where the group-level separation was clearer, the spatial distribution and extent of abnormalities differed considerably between individuals. Some healthy controls showed supra-threshold values, whereas some patients exhibited relatively limited abnormality evidence. These findings should not necessarily be interpreted as false-positive or false-negative lesion detection, because the Z-score maps represent model-derived diffusion abnormalities rather than histologically or clinically confirmed lesions. Their magnitude and spatial distribution may also be influenced by individual anatomical variability, diffusion measurement noise, tract reconstruction, spatial normalization, and variability within the healthy reference population. Accordingly, the individual maps should primarily be interpreted as visualizations of subject-specific abnormality evidence, whereas the systematic advantage of TractSpLearn is more appropriately assessed from the quantitative group-level analyses.

Taken together, the group-level and individual-level findings suggest that the patterns identified by TractSpLearn varied across cohorts and may depend on the consistency of patient-related imaging patterns. This interpretation is consistent with the stronger group-level differentiation and broader AK abnormality pattern observed in RHI, while the PPCS findings, including abnormalities detected in more patients but fewer affected tracts per abnormality-positive patient, suggest that the framework may also provide additional value in more heterogeneous patient populations, although to a lesser extent.

Several limitations should be considered though. First, the patient cohorts are relatively small, particularly for PPCS, which may limit the diversity and stability of the patient-related patterns learned by the model. Second, no independent lesion-level ground truth was available. Therefore, supra-threshold Z-score regions should be interpreted as model-derived abnormality evidence rather than confirmed pathological lesions, and the present study does not provide a formal assessment of diagnostic sensitivity or specificity. Third, the framework depends on several upstream processing steps, including diffusion model fitting, tract reconstruction, and spatial normalization. Although the same processing pipeline was applied to all subjects, methodological choices at these stages may introduce systematic, pipeline-dependent biases and thereby influence the sensitivity and spatial distribution of the detected abnormalities.  Finally, each diffusion parameter was modeled separately in the present study ,therefore, relationships between different parameters were not explicitly modeled. Future work could investigate whether joint modeling of multiple diffusion parameters, with a single multivariate model could better capture subtle and heterogeneous white-matter abnormalities.


\section{Conclusion}
Overall, TractSpLearn extends conventional tract-based normative modeling by incorporating patient-related information into individualized abnormality estimation. Rather than describing an unseen subject solely according to the magnitude of deviation from healthy controls, the proposed framework evaluates whether the observed tract profile shows greater preference for a patient-related representation than expected from the healthy reference distribution. TractSpLearn yielded positive findings in a higher proportion of patients than TractLearn in both cohorts, although positive findings in healthy controls also increased in the RHI cohort. Group-level differences were more pronounced in RHI than in PPCS, and the exploratory spatial analyses showed patterns that varied across cohorts, methods, and diffusion metrics. These findings support further evaluation of patient-informed modeling for the individualized characterization of white matter abnormalities. Because no independent lesion-level ground truth was available, the detected abnormalities should not be interpreted as confirmed lesions or as evidence of diagnostic accuracy; validation in independent cohorts and against clinical or biological measures is needed.

 \section {Data availability}
The dataset used in this study is not publicly available but may be obtained from the corresponding author at Lund University upon reasonable request, subject to applicable data access conditions.

\section {Code availability}
Tract segmentation was performed using TractSeg, available at \href{https://github.com/MIC-DKFZ/TractSeg}https://github.com/MIC-DKFZ/TractSeg. \\
The TractSpLearn code is available at \href{https://github.com/CyclotronResearchCentre/TractSpLearn}https://github.com/CyclotronResearchCentre/TractSpLearn.

\section*{Author Contributions}
Jiqing Huang-- Conceptualization; Methodology; Software (algorithm innovation and implementation); Formal analysis; Visualization; Writing – original draft.\\
Ali Al-Husseini-- Investigation (data acquisition for the repetitive head injuries cohort); Data curation; Resources.\\
Yi Chen-- Methodology (algorithm design); Writing – review \& editing.
Laurent Lamalle--  review \& editing.\\
Anna Gard-- Investigation (data acquisition for healthy controls and individuals with persistent post-concussive syndrome); Data curation; Resources.\\
Markus Nilsson-- Validation; Conceptualization (initial); Writing – review \& editing.\\
Mohamed Ali Bahri --Validation; Statistical analysis;Writing – review \& editing.\\
Niklas Marklund-- Resources (data provision); Protocol design;  Writing – review \& editing.\\
Christophe Phillips-- Conceptualization; Methodology; Project administration (algorithm and manuscript); Supervision; Writing – review \& editing.\\
Evgenios N. Kornaropoulos-- Conceptualization; Methodology; Supervision; Writing – review \& editing.\\
\section*{Acknowledgements}
This work was supported by the FRS-FNRS and computational resources from the CÉCI (Consortium des Équipements de Calcul Intensif, \url{https://www.ceci-hpc.be/} platform and the SEGI (Service Général de l’Informatique) at the University of Liège (ULiège). We are especially grateful to Prof. David Colignon for his assistance with resource allocation and technical support.
\printbibliography
\end{document}